\documentclass[twocolumn,prl,preprintnumbers,amsmath,amssymb,floatfix]{revtex4}

\usepackage{graphicx}
\usepackage{dcolumn}
\usepackage{bm}
\usepackage{verbatim}
\usepackage{tabularx}

\pacs{hey}

\def\beq{\begin{eqnarray}}
\def\eeq{\end{eqnarray}}
\def\bea{\begin{eqnarray}}
\def\eea{\end{eqnarray}}

\newcommand{\lsim}{ \mathop{}_{\textstyle \sim}^{\textstyle <} }

\newcommand{\gev}{\, {\rm GeV}}

\newcommand{\mev}{\, {\rm MeV}}

\def\oone{{\cal O}(1)}
\def\mpl{M_{Pl}}

\def\bibentry#1{\noindent\hangindent=2em #1}

\begin{document}

\preprint{CERN-PH-TH/2013-102}

\title{The Utility of Naturalness, and how its Application to \\
Quantum Electrodynamics envisages the Standard Model and Higgs Boson}

\author{James D. Wells}
\affiliation{
CERN Theoretical Physics, Geneva, Switzerland, and \\
Physics Department, University of Michigan, Ann Arbor, MI  USA }

\date{May 15, 2013}

\begin{abstract}
With the Higgs boson discovery and no new physics found at the LHC, confidence in Naturalness as a guiding principle for particle physics is under increased pressure. We wait to see if it proves its mettle in the LHC upgrades ahead, and beyond. In the meantime, in a series of ``realistic intellectual leaps" I present a justification {\it a posteriori} of the Naturalness criterion by suggesting that uncompromising application of the principle to quantum electrodynamics leads toward the Standard Model and Higgs boson without additional experimental input. Potential lessons for today and future theory building are commented upon.

\end{abstract}

\maketitle

\section{Introduction}

The discovery of the Higgs boson with mass of $126\gev$ (Aad et al.\ 2012; Chatrchyan et al.\ 2012) has been an exciting development in physics. At long last dynamics has been found that gives deeper insight into the origin of elementary particle masses. It has put to rest many ideas that were pursuing a different path, namely that the Higgs boson did not exist at all. Examples included approaches to technicolor and ``Higgsless theories." That chasm in the field is now resolved. 

A new chasm is coming to the fore. On one side are researchers convinced that the Higgs boson is unnatural all on its own. A supporting cast of particles and interactions are needed to protect it from destabilizing quantum corrections.  The so-called quadratic divergences in the Higgs boson self energy implies that quantum corrections are quadratically sensitive to the cutoff scale ($\Lambda$) of the theory: $\delta m_H^2\sim \Lambda^2$.  In this mindset it is unnatural for the Higgs boson mass to be lighter than the cutoff of the theory, which can be as high as $M_{Pl}$ ($\simeq 10^{18}\gev$). New physics must show up relatively light ($\Lambda\lsim {\rm TeV~scale}$) to keep these destabilizing influences in check. Candidate solutions are supersymmetry, composite Higgs, large or warped extra dimensions, and so on.

On the other side, some argue that there is no Naturalness problem with the Higgs boson mass being $126\gev$, and there is no need to posit any extra symmetries or principles to stabilize it there. They note that technically one can formulate the renormalizable theory with renormalized couplings and counter terms and order-by-order consistently assign parameters values that keep the Higgs mass light. In dimensional regularization, the most common technique to book-keep the infinities of the quantum field theory, there is no quadratic divergence explicitly manifested in the effective theory. Thus, as the argument goes, Naturalness is only a fuzzy philosophical notion, and should not be taken seriously.

The side one chooses on this question influences the research direction of the individual and the field. It is thus important to devote considerable reflection on the role of Naturalness in formulating quantum field theories. There is considerable history on Naturalness and related topics in the community, and the issue has become even more urgent in the context of the Higgs boson (Giudice 2008).  The central questions addressed here are whether Naturalness is a useful criterion by which to judge the value of a theory, and do efforts to replace unnatural theories with more natural theories lead to more fundamental insights into nature? I present here an {\it a posteriori} analysis of the utility of Naturalness in order to support a positive response to both questions.

\section{Absolute Naturalness}

Formulating the question of whether Naturalness is a useful concept suffers from imprecision if we do not define the term explicitly, yet broadly enough to capture all its needed usage. Let us begin with what we call Absolute Naturalness. {\it A theory has Absolute Naturalness if all dimensionless parameters and the ratio of any dimensionful parameters have explanation in terms of ${\cal O}(1)$ parameters.} A parameter possesses Absolute Naturalness with respect to the rest of the theory if it can be explained by ${\cal O}(1)$ parameters. There can be difference of opinion about the precise range of values $\oone$ implies, but when it is helpful to think precisely about the value let us say ${\cal O}(1)$, or its inverse, can be as low as $10^{-3}$.

An example of a parameter that is Absolutely Natural is $\Lambda_{QCD}$ in quantum chromodynamics. $\Lambda_{QCD}\sim 1\gev$ is Natural in this sense despite it being eighteen orders of magnitude smaller than the Planck scale ($M_{Pl}\sim10^{18}\gev$) because it can be explained by an ${\cal O}(1)$ parameter (strong coupling constant) near $M_{Pl}$, which upon renormalization group flow to the infrared diverges at $E\sim \Lambda_{QCD}$. Dimensional transmutation like this is a frequently used instrument in the theory toolbox when developing Natural theories. 

It is an implicit article of faith amongst most model-building theorists that Absolute Naturalness is a property of fundamental theories. The more Absolute Naturalness a theory has the more fundamental it is. Theories with very small numbers or large ratios are considered subject to further refinement. There are many examples in the literature of this, and practitioners know this well. Let us mention one. The Standard Model has Yukawa couplings that range from $10^{-6}$ for the electron to $\sim 1$ for the top quark. The large range of Yukawa couplings when confronted with the hypothesis of Absolute Naturalness suggests that the Standard Model is incomplete and new dynamics or symmetries can be identified that explain the fermion masses in a deeper theory possessing Absolute Naturalness. A comprehensive survey of theories of flavor, see for example Babu (2009), shows that the endeavor can be summarized as the attempt to explain large hierarchies of Yukawa couplings and mixing angles in terms of ${\cal O}(1)$ parameters, thereby constructing a theory with Absolute Naturalness. Whether the mechanism is radiative generation of small fermion masses or Froggatt-Nielsen field insertions, the new theory is burdened with much more complexity and parameters in order to satisfy the demands of Naturalness. 

The addition of many more fields, parameters and interactions is quite a price to pay for a mere philosophical premise -- especially one that leads to less simplicity in the theory. One can safely say that in the era of the Standard Model's ascendancy, the influence of simplicity and Ockham's razor to theory construction has paled in comparison to Naturalness.

\section{Technical Naturalness}

There is another notion of Naturalness that has been articulated from at least the early 1980's, which is called Technical Naturalness, or sometimes `t Hooft Naturalness (`t Hooft 1980). {\it A theory has Technical Naturalness even if some parameters are small, if an enhanced symmetry develops when the small parameter is taken to zero.} For example, a very light fermion mass is Technically Natural since an enhanced chiral symmetry emergences when the mass is taken to zero. In quantum field theory this protects the small parameter from any large quantum correction, and the small value is technically stable. 

Agreeing that theories with small parameters need to possess Technical Naturalness is less philosophically taxing than demanding nature must be described by a theory with Absolute Naturalness. Normally, Technical Naturalness is a short-hand descriptive statement, equivalent to saying that a small parameter in the effective theory is radiatively stable to a much higher cutoff scale.  It is the working hypothesis that fundamental theories should possess Absolute Naturalness, either explicitly or more often implicitly assumed in the field, that is subject to controversy and possible refutation upon further scrutiny. In the subsequent discussion the word Naturalness will be used to mean the stronger form of Absolute Naturalness, and Natural will be used to mean possessing the qualities of Absolute Naturalness.

\section{Justifying Naturalness}

Assuming  Naturalness as a law of nature imposes very strong constraints on model building. In the case of the Higgs boson, it leads not only to ideas like supersymmetry, which protects the Higgs boson from having a large mass, but the devotion one has to strict Naturalness leads to radically different hypothesized superpartner spectra. Compare the spectrum of heavy superpartners in PeV scale (Wells 2003; Wells 2004) or split supersymmetry (Arkani-Hamed \& Dimopoulos 2004; Giudice \& Romanino 2004), where standard views of Naturalness are not so strictly bowed to, versus the spectrum of supersymmetry that requires no parameter finetunings of more than one percent (Ross et al.\ 2012).

In the past, Technical Naturalness has been used to understand experimental results that have already been measured. For example, the masses of the pions, proton and neutron are understood well from symmetries and Technical Naturalness. We know from asymptotic freedom of quantum chromodynamics (QCD) that the perturbative gauge coupling in the ultraviolet flows to strong value at the low scale and confinement happens at $\Lambda_{QCD}\sim 1\gev$. This gives the characteristic scale of the hadrons in the theory, and the proton and neutron obtain mass approximately equal to this scale. However, the pion masses are much lower, and can be understood as the Goldstone bosons of $SU(2)_L\times SU(2)_R\to SU(2)_V$ flavor symmetry breaking. The mass is exactly zero when there are no explicit quark masses in the theory, and this ``hierarchy" is very well understood. Furthermore, few are distressed that the proton mass $m_p\sim 1\gev$ is much less than the Planck mass $\mpl\sim 10^{19}\gev=(G_N)^{-1/2}$. The reason is as stated earlier: an ${\cal O}(1)$ number, namely the QCD gauge coupling, is an input at some high scale that through renormalization group flow generates an exponentially suppressed scale through dimensional transmutation. This is not concerning because no very big or very small numbers were needed as input. In other words, there is an explanatory theory that possesses Absolute Naturalness.

However, there is an unresolved Naturalness problem in QCD. Namely, Goldstone bosons are not exactly massless, but gain small mass due to explicit breaking from quark masses. These quark masses are neither $\mpl$ nor $\Lambda_{QCD}$. They are very small compared to both. The up and down quark masses are roughly $m_q\sim 10\mev$. This is about a $10^{-2}$ suppression with respect to $\Lambda_{QCD}$, which is borderline acceptable from the Naturalness point of view, even though there is no obvious connection at first sight between $\Lambda_{QCD}$ and quark masses. The light quark masses are more than a factor of $10^{-20}$ suppressed compared to $\mpl$, which of course looks much worse, especially since, from the point of view of QCD, the theory is vectorlike, and there is no obstacle to giving the quarks much higher mass. The theory is very obviously not Natural.

What we have seen here is that Absolute Naturalness is satisfied when explaining the proton and neutron mass. However, if we want Absolute Naturalness to explain the small non-zero pion masses, or equivalently the light quark masses, we must go to a deeper theory. Writing down small, explicit masses is not allowed in a theory with Absolute Naturalness without further explanation.

\section{Quantum Electrodynamics and Naturalness}

We could proceed further with a discussion about applying Absolute Naturalness demands on the quarks of QCD, but quantum electrodynamics (QED) provides exactly the same conundrum, except that it is simpler to discuss and the problem to overcome is numerically more severe. 

Let us consider the QED theory as it was known in its early days. The lagrangian is quite simple:
\beq
{\cal L}=\frac{1}{4}F_{\mu\nu}F^{\mu\nu}+i\bar \psi \gamma^\mu (\partial_\mu-ieA_\mu)\psi+m_e\bar \psi\psi
\eeq
where $\psi$ is the spin-1/2 electron field, $A_\mu$ is the spin-1 photon field, $e\simeq 0.31$ is the gauge coupling constant and $m_e=0.511\mev$ is the electron mass. The coupling constant appears to satisfy the demands of Absolute Naturalness, but the electron mass does not.  

One violation of the electron mass with respect to Absolute Naturalness is that $m_e/\mpl\simeq 10^{-23}$. However, one might object that this involves the mysteries of gravity which are too difficult to sort out and that perhaps all of the Naturalness considerations of the Standard Model can be satisfied at mass scales far removed from $\mpl$. This attitude of ignoring the Planck scale is suspect, and a theory that truly possesses Absolute Naturalness should have no unexplained large hierarchies, including with respect to $\mpl$. However, this discussion can be circumvented, since the problem with the electron mass is already transparent without invoking $\mpl$.

In 1933/34 Fermi proposed his theory of nuclear beta decay, which in his mathematical language was governed by the interaction (Fermi 1934)
\beq
H=g\, \{ Q\psi(x)\varphi(x)+Q^* \psi^*(x)\varphi^*(x)\},
\eeq
where $Q$ is a proton-neutron transition operator, and $\psi$ and $\varphi$ represent the electron and neutrino. Fermi used data known at the time to estimate the dimensionful coupling constant $g$ to be
\beq
g=4\times 10^{-50}\, {\rm cm}^3\, {\rm erg} ~ (=3.25\times 10^{-6}\gev^{-2}).
\eeq
This implies a Fermi scale of $M'_F=g^{-1/2}=555\gev$. Thus, in 1934 it was explicitly in the literature that a scale that governed the interactions of electrons was six orders of magnitude greater than the electron mass itself.  

In modern language the Fermi interaction theory is written as  the four fermion operator
\beq
\frac{G_F}{\sqrt{2}}\, \left[ \bar u\gamma^\mu\frac{(1-\gamma^5)}{2} d\right]\, 
\left[\bar e\gamma_\mu \frac{(1-\gamma^5)}{2} \nu\right],
\eeq
where $G_F=1.17\times 10^{-5}\gev^{-2}$ is the Fermi constant.  This constant defines a new mass scale $M_F=(G_F)^{-1/2}=293\gev$.  The violations of Absolute Naturalness can now be phrased as a ratio that is unnaturally too small: 
\beq
\frac{m_e}{M_F}=1.7\times 10^{-6}~~{\rm (Absolute~Naturalness~problem)}
\eeq
If pursuing Absolute Naturalness is a valid guide to constructing more fundamental theories, one should be able to apply the principle to this problem and see that it could have guided one to deeper insights if adhered to uncompromisingly.

It should be reiterated parenthetically that in today's language and knowledge we know that small $m_e$ is Technically Natural, as a chiral symmetry enhancement develops in the limit of $m_e\to 0$. The small mass value is therefore stable to quantum corrections of the theory, which was understood in the early days of quantum electrodynamics (Muryama 2000), even though they did not yet have efficient language and theory constructs to simply justify its necessity.  Nevertheless, Absolute Naturalness says that $m_e\bar\psi\psi$ really should be $M_F\bar\psi\psi$. 

\section{Realistic Intellectual Leaps to the Standard Model}

At this point we have come to the conclusion that the electron mass is unnaturally small in QED. But what to do about it? In this section I will outline a series of Realistic Intellectual Leaps (RILs) that I think would be achieved in short order, and were ripe to do so at the time, if only researchers were committed to Naturalness and were unwaveringly tenacious at trying to explain the smallness of the electron mass.  

At times I will put forward historical context that is meant to support the corresponding RIL. This is not intended to be a rigorous or comprehensive history, but rather anecdotes showing that many of the RILs were at hand even during the early days of QED. Nevertheless, the argument here is not reliant on historical analysis, but of conceptual inevitabilities once Naturalness is declared a primary goal. Let us begin with the first RIL that would set it all in motion.

{\bf RIL \#1}: {\it The value of $m_e$ is closer to zero than it is to $M_F$.}

On the surface this may look like a strange statement, but we hear such statements frequently in many different contexts, and they have been part of physics discussions time immemorial.  From the earliest days of Newtonian mechanics when considering orbits of ``massless" planets, to present day collider physics where $b$ quark is ``massless" in comparison to the characteristic energies of the collisions, researchers have realized that for all practical purposes objects sometimes have a mass ``closer to zero" than the other characteristic scales or masses of the problem.

What this means operationally for the electron mass in QED theory, and the realistic intellectual leap implied here, is to pursue a path where a way is somehow provided to forbid the electron mass at zeroth order in the theory, and then somehow reinstate a small correction to zero by some other means. I consider this the biggest intellectual leap of all the ones that I will present. Nevertheless, the fact that I am articulating this approach now is evidence that the leap is possible, and my understanding derives from the collective wisdom of a large number of researchers who have been pursuing this path for years in many different guises, most importantly in flavor physics. There has been no experimental discovery since the 1950's that has forced this thinking on us, but from pure thought banishing masses ``at tree level", or ``at the renormalizable level", or ``zeroth order" has been deduced as a worthy starting point for explanation. And it was devised quickly in modern times upon the ascendancy of Naturalness as a criterion to be satisfied when explaining parameter hierarchies. Therefore, there is no reason to discount the possibility that researchers would have had these obvious and similar thoughts if only Naturalness in QED were pursued as a paramount virtue.

Inspection of the QED lagrangian tells us immediately that the problem is the existence of the gauge-invariant and Lorentz invariant operator $\bar \psi\psi$.  This mass operator is dimension three and so to round out the dimensions to four one needs a massive coupling, which Absolute Naturalness demands should be similar to $M_F$. A reasonable starting path is to somehow recast the theory to make $\bar\psi\psi$ not an invariant, in the spirit of recognizing that $m_e$ is ``closer to zero than it is to $M_F$."

If adherence to Absolute Naturalness is our primary concern, no ugliness or complexity should stop us from finding a way to banish this offending operator $\bar\psi\psi$. In time it would be inevitable that theorists would recognize that there is something special about that operator compared to others. This leads to the next RIL.

{\bf RIL \#2}: {\it The representation structure of the Lorentz group allows us to write QED in two-component spinors, which is seen to demonstrate a qualitatively unique feature to the mass operator.}

In two-component notation the QED lagrangian is 
\bea
{\cal L}&=&\frac{1}{4}F_{\mu\nu}F^{\mu\nu}+i\psi^\dagger_L \gamma^\mu (\partial_\mu-ieA_\mu)\psi_L \\
& & +\psi^\dagger_R \gamma^\mu (\partial_\mu-ieA_\mu)\psi_R +m_e(\psi^\dagger_L\psi_R+\psi^\dagger_R\psi_L).
\nonumber
\eea
We see that the mass term mixes the right and left handed components of the spinor, whereas the kinetic term and gauge interactions do not.  

RIL \#2 is certainly inevitable for anyone who thinks semi-seriously about spinor representations and the invariant operators that can be written from them. Furthermore, we can survey some historical anecdotes to show that this understanding was immediately at reach even before QED was known.

In the 1930's and 1940's researchers may not have recognized the qualitative difference of the mass term compared to others in the language that we describe above. However, it was very well known that the representation components of the spinor interacted with the mass operator differently than to the kinetic term and gauge potential term in the Dirac equation. For example, van der Waerden (1929) introduced two-component spinor notation, and Laporte and Uhlenbeck (1931) popularized it in an influential article very soon thereafter.

At the time, equations of motion were more fashionable to write down than lagrangians, but the insight of RIL \#2 is just as plausible to infer. For example, Laporte and Uhlenbeck (1931) split the Dirac equation in two parts consistent with two-component notation and wrote 
\bea
mc\chi_\ell - \left( \frac{h}{i}\partial^{\dot\sigma}_{~\ell}+\phi^{\dot\sigma}_{~\ell}\right) \psi_{\dot\sigma} & = & 0 \\
mc\psi_{\dot m}+\left( \frac{h}{i}\partial_{\dot m}^{~\lambda}+\phi_{\dot m}^{~\lambda}\right) \chi_\lambda & = & 0,
\eea
where the dotted and undotted parameters index the $(\frac{1}{2},0)$ and $(0,\frac{1}{2})$ spinor representations of the Lorentz group decomposed in $SU(2)\times SU(2)$. The $\phi$ field is the spinor-represented gauge field of electromagentism.
It is straightforward to see from this first equation that the mass term interacts with the undotted indices and the kinetic and potential term with the dotted, and vice-versa for the next equation. Thus, there is a qualitative difference in the Lorentz structure of the mass term compared to the kinetic and gauge potential interaction terms.

Now that we have established this unique Lorentz structure of the mass term, the next leap is to recognize that the $\bar\psi\psi$ term can now be banished.

{\bf RIL \#3}: {\it The electron mass can be forbidden by assigning $\psi_L$ different properties than $\psi_R$, thereby disallowing $\psi^\dagger_L\psi_R+\psi^\dagger_R\psi_L$ as an invariant of the theory.}

The first thought would be to give $\psi_L$ and $\psi_R$ different electric charge, but that violates experiment badly since there are not two different electric charges to be seen. 
The electric charge for both we must keep at $-1$, but we can assign different charges for each under a new symmetry $G$.  A simple concrete start to this would be to let $G$ be a new abelian group $U(1)'$, in analogy with the $U(1)_{EM}$ group of QED. We need only give $\psi_R$ a different charge under $U(1)'$ than $\psi_L$. One simple choice is to  assign $\psi_R$ double the charge of $\psi_L$. In other words, our spectrum is
\bea
& {\rm Under~}U(1)_{EM}\times U(1)': & \nonumber \\
&\psi_L=(-1,-1),~~\psi_R=(-1,-2)&
\eea
With these charge assignments the $\bar \psi\psi$ term is no longer allowed.

The next step is to somehow regain the electron mass through some other means. Everything would be tried, and in time researchers would be led to a candidate solution.

{\bf RIL \#4}: {\it The electron mass can be reintroduced by adding a condensing scalar field $\Phi$ with charges $(0,1)$ under $U(1)_{EM}\times U(1)'$ that allows the interaction $y_e \psi^\dagger_L\Phi\psi_R+h.c.$, which leads to the electron mass $m_e=y_e\langle\Phi\rangle$.}

This may look to be extraordinarily optimistic that scientists with 1950's-limited experimental knowledge could have come to this insight. However, such ideas were already firmly in the air, which gives further support to the notion that pursuers of QED Naturalness can have this insight without it in hindsight seeming too radical or too genius. Ginzburg and Landau's 1950 paper ``On the Theory of Superconductivity" (Ginzburg \& Landau 1950) already had the seeds of this kind of solution.  It was in this paper, building on Landau's 1937 mean-field-theory account of phase transitions (Landau 1937), that researchers read about a complex scalar function $\Psi$. A power series functional expansion of $\Psi$ was presented that looks very similar to the Higgs potential of today, and its minimization enables $|\Psi|^2$ to obtain a non-zero value (vacuum expectation value) below a critical temperature.   Furthermore, $\Psi$ was recognized to have a phase symmetry, equivalent to transformations under $U(1)_{EM}$, and the energy density of the system contained the term
\beq
\frac{1}{2m}\left| -i\hbar\, {\rm grad}\, \Psi -\frac{e}{c}{\bf A}\Psi\right|^2
\label{eq:covariant}
\eeq
where it was noted that gauge invariance required replacing $i\hbar\,{\rm grad}$ to $i\hbar\,{\rm grad}-(e/c){\bf A}$, which is what we today call the covariant derivative.

What followed in Ginzburg-Landau's theory of superconductivity in 1952 was an expansion of the free energy in terms of the relevant quantities, including interactions of the $\Psi$ field with the photon field. This was done in the context of pure electromagnetism without explicit fermion introductions; however, the methodology of introducing a scalar function that transforms under a phase symmetry, has non-zero vacuum expectation value and couples to all the degrees of freedom of the theory would not escape researchers' attention. The work by Ginzburg and Landau was recognized widely and immediately as important in the community, as evidenced, for example, by its mention in Shoenberg's influential 1952 book on superconductivity (Shoenberg 1952), wherein Shoenberg declares that Ginzburg and Landau's recent theory treats superconductivty ``in a more fundamental way" than Londons' theory.

In short, the inevitable steps are to introduce a $\Phi$ field of the appropriate charge, couple it to the electron as above, and  enhance the Lagrangian to include kinetic terms and potential for the scalar field
\beq
\Delta {\cal L}=|\partial_\mu\Phi|^2-\mu^2\Phi^\dagger \Phi-\lambda(\Phi^\dagger\Phi)^2.
\eeq
Once this step occurs the Higgs mechanism follows almost immediately. A systematic study of the parameters of this potential would easily recognize that when $\mu^2<0$ there is a vacuum expectation value of the Higgs boson and stability of the potential requires $\lambda>0$ at the renormalizable level.  This leads to a minimum of the potential where the $\Phi$ field has a vacuum expectation value of $\langle \Phi\rangle=-\mu^2/\lambda$.

Once the Higgs mechanism is understood progress would be very fast. It would be recognized that the fermion mass is obtained by $m_e=y_e\langle \Phi\rangle$, completing the claim of RIL \#4. Thus, the electron mass is generated, and for suitable choice of $y_e$ and $\langle \Phi\rangle$ the correct mass is obtained.

At this point there is a lot of algebra and analysis that even beginning students can do to explore the consequences of the theory introduced. It would be seen very quickly that if the new $U(1)'$ were a global symmetry it would mean that a new massless boson (Goldstone boson in our modern language) arises, which would be understood to be in conflict with experimental results. However, rather than considering $U(1)'$ to be a global symmetry, there is a more obvious thing to do, which leads to our next RIL.

{\bf RIL \#5}: {\it To avoid massless states in conflict with experiment, the new symmetry $U(1)'$ should be a local gauge symmetry analogous to $U(1)_{EM}$. When $U(1)'$ is broken by $\langle\Phi\rangle$ the new photon gets a mass, $m_{A'}^2=\frac{1}{2}g'^2\langle\Phi\rangle^2$ and there remains a physical propagating scalar boson with mass $m_\varphi^2=2\lambda\langle\Phi\rangle^2$.}

RIL \#5 is just a consequence of trivial calculations, and it is hardly an intellectual leap given the assumption that the first four RILs have been achieved. The implications are huge, however. What is being predicted is a new neutral massive gauge boson that interacts with the electron in a chiral way, and a new scalar boson that interacts with the electrons. This is a qualitative conclusion about nature that we know today is correct (the $Z$ boson and Higgs boson of the Standard Model), but would have likely been met with significant skepticism by the community. This is the first RIL that has falsifiable consequences, and puts the theory at risk.  It is here where the program we followed that started with Naturalness, which is more ``philosophical" rather than ``scientific", has led to firm ``scientific" predictions and is subject to verification or refutation over time, in principle.

We are not done. As it stands our theory is sick, because there are anomalies.   The $U(1)'$-graviton-graviton and $U(1)'^3$ anomaly cancelation conditions are not met among the fermions. Researchers carrying out full quantum calculations of the theory, which they are wont to do, would inevitably find inconsistent results unless various sums of the charges of the fermions are zero. This needed cancellation of charges does not happen in our theory as it is presently constituted, and the anomaly persists. This necessitates additional fermions in the spectrum that are charged under $U(1)'$.  In other words, we have argued for a new RIL.

{\bf RIL \#6}: {\it Researchers trying to analyze the theory at the quantum level would discover that $U(1)'$ has chiral anomalies with respect to the electron charges, and would be forced to add canceling exotic fermions.}

There are many possibilities to cancel the anomalies in this augmented quantum electrodynamics theory of the electron and photon. This is not a difficult diophantine equations exercise to pursue, and a supervised undergraduate thesis could list a large number of possibilities. One of the entries of such a compilation would be the following set of exotic fermions:
\beq
{\rm Exotics:~}6Q'_{1/3}+3Q'_{4/3}+3Q'_{-2/3}+1Q'_{-1}
\eeq
where $nQ'_q$ means $n$ copies of fermions with charge $q$. They add to our original fermions $1Q'_{-1}+1Q'_{-2}$.
These charge assignments and multiplicities are exactly those of the Standard Model fermions under twice hypercharge (see, e.g., table 1 of Wells (2009)).

There was infinite freedom in how I chose $U(1)'$ charge assignments for $\psi_L$ and $\psi_R$, and the magic of recovering the Standard Model spectrum was that I chose the simple case of the $\psi_R$ charge being twice that of $\psi_L$. We did not have to choose those charges, of course. Furthermore, we could have even made $G$ a non-abelian group, as that would do just as well in protecting the fermion masses. For example,  $\psi_L$ could transform as a fundamental under $SU(N)$, $\psi_R$ as a singlet, and $\Phi$ as an anti-fundamental.  Nevertheless, a key qualitative point remains in all choices, and that is that exotics are expected in the spectrum.  Furthermore, the set of exotics that satisfies the anomaly cancelation conditions can be repeated many times and still all the constraints are satisfied. This we can call the number of generations.

\section{Objections to the Resulting Theory}

The reader may object at our exotically augmented QED theory and say that in our attempt to solve the Naturalness problem of the electron we have introduced several other problems. To start with, we have introduced more parameters, more fields, and more complexity in the theory, all in the service of a debatable philosophical proposition. However, the status of these objections are also based on debatable philosophical notions of simplicity and economy, and if we posit that Naturalness has higher status, then our result carries through.

A more worrisome objection is that we have replaced one problem with Absolute Naturalness (the electron mass) for another Naturalness problem, and so have not achieved our objective. If the $\Phi$ mass is $-\mu^2\sim m_F^2$ (natural) then $y_e\sim 10^{-6}$ (unnatural), or phrased differently, if $-\mu^2\sim m_e^2$ (natural) then $-\mu^2/M_F^2\sim 10^{-6}$ (unnatural). Either way there is a problem with respect to Absolute Naturalness. Thus, all we have done is recast the Naturalness problem rather than solved it.

There are several ways to approach this objection. However, the one I would like to advocate is that it means simply that we are not yet done -- the problem is indeed not yet solved completely. Some researchers can scrap a commitment to scalars as introduced above, and start afresh with a new idea to tackle the electron mass problem. But what new direction to take? And, in any event, we know now they would be wrong. Or, researchers can attack with vigor the new problem of scalar boson mass Naturalness in hopes that a solution presents itself to that recasting of the problem.  Over time, ideas of supersymmetry, extra dimensions and landscape statistics arguments may all surface, as indeed they have in the real history, to solve the problem.  

The key point here is that directly attacking the electron mass Naturalness problem in quantum electrodynamics inevitably presents to us a discrete path with a narrow gateway that is characterized by all the features we discussed above (heavy gauge boson, propagating scalar, exotic fermions). Upon passing through that gateway many other subsequent gateways might present themselves to us as well to solve the next problems of Naturalness (supersymmetry, extra dimensions, etc.), which may also have recasting objections at their next level. For example, one objection to supersymmetry is the $\mu$ problem which requires that an explicit mass in the superpotential for vectorlike Higgsino interactions be near the weak scale rather than the more natural Planck scale. There are proposed solutions. An objection to large extra dimensions is that it recasts the large hierarchy of weak scale to Planck scale as a large hierarchy of the weak scale to the compactification scale. There are proposed solutions to this. And so on.

The need to continue on the quest toward a fully Natural theory does not diminish the uniqueness, validity and insight of our first gateway.  It is likely that we will need to find many such gateways on the path to a theory of full Absolute Naturalness.

\section{Conclusions}

One approach to justifying Absolute Naturalness as a guide to theory model building is to consider how science could have progressed in the past if researchers had firmly devoted themselves to the principle. Would it have led us astray or would it have led toward more fundamental theories? Although it is conceivable that it could lead us astray at times, I have presented here evidence of its salutary influence. 

In this article I have considered QED as a theory in gross violation of Absolute Naturalness, and followed plausible steps scientists could have taken if they were wholly devoted to constructing a theory without this problem. The inferences that result are 1) the existence of an extra scalar Higgs boson field that couples according to the mass of the electron, 2) an exotic gauge symmetry with a massive gauge boson, 3) parity violation in the fermion interactions with the gauge boson, 4) the necessity of additional exotic fermions to cancel anomalies. These inferences are necessitated by the approach we took. What's also possible is  5) the prospect of multiple copies (generations) of exotic fermions, and 6) the prospect of non-abelian gauge symmetries chiral-protecting the fermion masses.

The steps along the way would have required creativity and dedication to discovery, both at plausible levels, and also would have required strength in the face of criticisms regarding lack of simplicity and the complexity invoked, and the speculations of new particles and new forces. We may be finding ourselves in a similar situation today with respect to the continuing but pressure-strained research to forge a theory of particle physics possessing Absolute Naturalness, which invariably predicts new dynamics, new principles and exotic new particles.

The principle of Naturalness cannot be derived from first principles, and its invocation in science is more of a product of intuition against the likelihood of large numbers conspiring together to give small numbers than it is on rigorous deduction. Nevertheless, the concept bears fruit and is satisfied with respect to our QED example here and other examples to be found in the literature. It remains to be seen if the hard pursuit of Naturalness for the Higgs boson sector has correctly predicted new physics, yet to be seen, that is near the Higgs boson mass scale. 

\medskip\noindent
{\it Acknowledgments:} I am grateful for discussions with R. Akhoury, D. Dieks, J. van Dongen, G. Giudice, G. `t Hooft,  S. Martin and C. Smeenk. 

\bigskip
\noindent
{\bf References}\medskip

\bibentry{Aad, G.\ et al.\ (ATLAS Collaboration). (2012).\ Observation of a New Particle in the Search for the Standard Model Higgs Boson with the ATLAS Detector at the LHC. {\it Phys.\ Lett.}\ B716, 1-26 [arXiv:1207.7214].}

\bibentry{Arkani-Hamed, N.\ \& Dimopoulos, S. (2004).\ Supersymmetry unification without low energy supersymmetry and signatures for fine-tuning at the LHC. {\it JHEP} 0506, 073 (2005) [hep-th/0405159].}

\bibentry{Babu, K.S. (2009).\ TASI Lectures on Flavor Physics. arXiv:0910.2948 [hep-ph].}

\bibentry{Chatrchyan, S.\ et al.\ (CMS Collaboration). (2012).\ Observation of a new boson at a mass of 125 GeV with the CMS experiment at the LHC. {\it Phys.\ Lett.}\ B716, 30-61 [arXiv:1207.7235].}

\bibentry{Fermi, E. (1934).\ Versuch einer Theorie der $\beta$-Strahlen. I. {\it Zeit.\ f\"ur Phys.}\ 88, 161.}

\bibentry{Ginzburg, V.L.\ \& Landau, L.D. (1950).\ On the Theory of Superconductivity. {\it JETP} 20, 1064. For English translation see Landau (1965).}

\bibentry{Giudice, G. (2008).\ Naturally Speaking: The Naturalness Criterion and Physics at the LHC. In G. Kane, A. Pierce (eds.), {\it Perspectives on LHC Physics}, 2008 [arXiv:0801.2562].}
  
\bibentry{Giudice, G.\ \& Romanino, A.\ (2004).\ Split supersymmetry. {\it Nucl.\ Phys.}\ B699, 65 (2004) [hep-ph/0406088].}

\bibentry{`t Hooft, G. (1980).\ In {\it Recent Developments in Gauge Theory.} New York: Plenum Press.}

\bibentry{Landau, L.D. (1937).\ On the Theory of Phase Transitions. I: {\it Phys.\ Z.\ Sow.}\ 11, 26. II: {\it Phys.\ Z.\ Sow.}\ 11, 545. For English translation see Landau (1965).}

\bibentry{Landau, L.D. (1965).\ {\it Collected Papers of L.D. Landau.} ed.\ and intro.\ by D.\ ter Haar. New York: Gordon and Breach.}

\bibentry{Laporte, O.\ \& Uhlenbeck, G.E. (1931).\ Application of Spinor Analysis to the Maxwell and Dirac Equations. {\it Phys.\ Rev.}\ 37, 1380.}

\bibentry{Murayama, H. (2000).\ ``Positron Analogue" in Supersymmetry Phenomenology. {\it Proceedings of 1999 ICTP Summer School.} arXiv:hep-ph/0002232.}

\bibentry{Ross, G.G., Schmidt-Hoberg, K.\ \& Staub, F. (2012).\ The Generalised NMSSM at One Loop: fine Tuning and Phenomenology. {\it JHEP} 1208, 074 (2012) [arXiv:1205.1509].}

\bibentry{Shoenberg, D. (1952).\ {\it Superconductivity.} 2nd ed. Cambridge: Cambridge University Press.}

\bibentry{van der Waerden, B. (1929).\ In {\it G\"ottinger Nachrichten}, page 100.}

\bibentry{Wells, J.D. (2003).\ Implications of supersymmetry breaking with a little hierarchy between gauginos and scalars. hep-ph/0306127 (SUSY 2003, Tucson).}

\bibentry{Wells, J.D. (2004).\ PeV-Scale Supersymmetry. {\it Phys.\ Rev.}\ D71, 015013 (2005) [hep-ph/0411041].}

\bibentry{Wells, J.D. (2009).\ Lectures on Higgs Boson Physics in the Standard Model and Beyond. {\it British Universities Summer School of Theoretical Particle Physics}, University of Cambridge and University of Liverpool [arXiv:0909.4541].}

\end{document}